\documentclass[3p,times]{elsarticle}
\usepackage{graphics}
\usepackage{eurosym}
\usepackage{amsfonts}
\usepackage{amssymb}
\usepackage{amsmath}
\usepackage{booktabs}
\usepackage{xcolor}

\usepackage{graphicx}
\usepackage{graphicx,epsfig}
\usepackage[font={footnotesize,it}]{caption}
\usepackage[colorlinks=false,
linkcolor=red,
urlcolor=red,
citecolor=blue]{hyperref}

\usepackage{float}
\usepackage{color}
\usepackage{mathptmx}
\usepackage{anyfontsize}
\usepackage{t1enc}
\usepackage{amssymb}

\usepackage[figuresright]{rotating}

\begin{document}

\begin{frontmatter}




\title{Pulsar PSR B0943+10 as an isotropic Vaidya-Tikekar type compact star : A comprehensive study}


\author{Jitendra Kumar}
\ead{jitendark@gmail.com}
\author{Puja Bharti}
\ead{pujabharti06@gmail.com}

\address{Department of Applied Mathematics, Central University of Jharkhand, Ranchi-835205, India}

\begin{abstract}
In this paper, we have constructed a model for well behaved isotropic compact star in the presence of charged perfect fluid, by considering a static and spherically symmetric metric in Schwarzschild's canonical coordinate system. To put the resulting differential equations into a closed system, we have employed the Vaidya $\&$ Tikekar (J. Astrophys. Astron. \textbf{3}:325, 1982) form of the metric potential $g_{rr}$. The resulting energy-momentum components, i.e., energy-density and pressure contain six constants; two of these are determined through the junction condition (matching the interior with the exterior Schwarzschild solution) and by the property of vanishing pressure on the boundary. The remaining constants are constrained by requirements of a real compact star. The physical acceptability of our model is tested using the data of the pulsar PSR B0943+10. Using graphical analysis and tabular information we have shown that our model obeys all the physical requirements. The stability of this model is evaluated using the Tolman-Oppenheimer-Volkoff equation, the adiabatic index and the Harrison-Zeldovich-Novikov Criterion and it has passed the  evaluation.
\end{abstract}


\end{frontmatter}


\section{Introduction}
	In the general theory of relativity, the Einstein's field equations establish a relation between the geometry of space-time and the distribution of matter within it.  It has been a compelling field for both mathematicians and physicists to discover new analytical solutions for these equations. The very first exact solution of Einstein's field equations representing a bounded matter distribution was provided by Schwarzschild \cite{child}. This encouraged the researchers to search for physically applicable solutions for the same. It is still an  interesting topic to researchers these days. The exact solution plays a crucial role within the development of varied areas of a gravitational field like black hole solution, gravitational collapse, solar system test, modelling of pulsars like PSR B0943+10 and so on.

Pulsar PSR B0943+10 \cite{psr} (discovered at Pushchino in December 1968), is 2000 light-years from Earth. The pulsar is about 5 million years old, which is relatively older for a pulsar. PSR B0943+10 is one of the brightest pulsars at low frequency \cite{bilous}. It has an interesting characteristic that it emits both radio waves and X-rays. In the radio band, PSR B0943+10 is one of the most studied pulsars showing the mode-switching phenomenon. In fact, at irregular intervals, every few hours or less, PSR B0943+10 switches between a radio bright mode with highly organized pulsations and a quieter mode with a rather chaotic temporal structure \cite{suleimanova}. Its 0.1–10 GHz spectrum is very steep \cite{malofeev}. PSR B0943+10 exhibits a very interesting behavior in the X-ray band as well. It was the first rotation-powered pulsar exhibiting variations in its X-ray emission \cite{meregh}, contradicting the common view that rotation powered neutron stars are source of constant X-ray emission.

Due to the robust  nonlinearity of Einstein’s field equations and thus the shortage of a comprehensive rule to get all solutions, it becomes difficult to locate any new exact solution. Thousands of exact solution of the field equations describing an outsized number of stellar objects varying between perfect fluids, charged bodies, anisotropic matter distributions, higher-dimensional stars, exotic matter configurations, etc. are present so far. But most of them are physically irrelevant within the relativistic structure of compact stellar objects. For obtaining exact solutions describing static compact objects, some impositions like space-time dimensionality, symmetry requirements, an equation of state relating the pressure and energy density of the stellar fluid, the behavior of the pressure anisotropy or isotropy, vanishing of the Weyl stresses are made to make the problem mathematically more identifiable \cite{maur17}. 

Though there is a good range of stellar solutions exhibiting deviation from sphericity, spherical symmetry is the closest natural assumption to elucidate stellar objects. There exists a collection of static, spherically symmetric solutions which provides useful guide to the literature \cite[see][]{lake, phani}. In this regard, the primary model had been proposed by Tolman \cite{tov1}, which was followed by some generalizations made by Wyman \cite{man}, Leibovitz \cite{vitz}, and Whitman \cite{hitman}. Bayin \cite{bayin} then used the strategy of quadratures and gave new astrophysical solutions for the static fluid spheres. The study of Sharma et al. \cite{sharma} and Ivanov \cite{ivanov} shows that the presence of an electrical field affects the values of surface redshifts, luminosities, and maximum mass of compact objects. Ray et al. \cite{ray} performed the charged generalization of Bayin \cite{bayin}. Mak $\&$ Harko \cite{mak} and Komathiraj $\&$ Maharaj \cite{komaA, komaB} highlighted the particular incontrovertible fact that the electromagnetic field features an important role in describing the gravitational behavior of stars composed of quark matter. Models constructed in this manner are proven to be useful in describing the physical properties of compact relativistic objects with different matter distributions. There are several investigations on the Einstein-Maxwell system of equations for static charged spherically symmetric gravitational fields for eg. \cite{kiess, fatema, murad}.

Exact solutions of the field equations for various \lq neutral as well as charged static spherically symmetric configurations' for anisotropic pressure compatible to compact stellar modeling have been obtained in numerous works \cite[see][]{her85, her97, her01, mak02, mak2002, mak03, her08, maha12, maurya12, maurya13, maurya14, pantA, pantB}. In recent times various models of relativistic stars have been found with anisotropic pressures \cite[see][]{bhar, iva17, maurya18, prasad, das}. However, in addition to this, it is necessary to keep in mind the compact stars with isotropic pressure, as at times this may be typically thought to be the equilibrium state of gravitating matter. Physical analysis indicates that isotropic models may even be accustomed to describe compact charged spheres. Some samples of isotropic stars with an electromagnetic field are often seen in \cite{thiru06, komathi07, maha07, thiru09}. Various comprehensive investigations of charged isotropic spheres are contained among the works \cite{ivanov, kiess, fatema, murad, matondo, kumar1}. 

There is not enough information regarding the equation of state of matter contained within the interior of compact stars. This makes it difficult to apply analytic solutions to the equation of relativistic stellar structure to acquire its understanding \cite{latti2001,latti2004}. Oppenheimer $\&$ Volkof \cite{tov2} technique and Tolman \cite{tov1} method are two customary methodologies which are generally followed to acquire a realistic stellar model. In the first approach, we start with an explicit equation of state. The integration starts at the center of the star with a prescribed central pressure and iterated till the surface of the star has been reached, i.e., where pressure diminishes to zero. Normally such input equations of state does not yield closed-form solutions. In the second approach, Einstein’s gravitational field equations need to be solved. For a static isotropic perfect fluid case, the field equations can be reduced to a set of three coupled ordinary differential equations in four unknowns. After getting exact solutions, one can solve the field equations by considering one of the metric functions or the energy density as priori. Consequently, the equation of state can be computed from the resulting metric. Since non-physical pressure-density configurations are found more frequently than physical ones, a new solution that ought to be regular, well behaved, and can reasonably model a compact astrophysical stellar object is always appreciated \cite{md.}.
We are going to follow Tolman's methodology in this paper and specify one of the gravitational potentials as the Vaidya $\&$ Tikekar \cite{vaidya} potential which has been shown to model superdense stars in several papers. 

The presence of five unknown functions and only three basic field equations permits one to specify the metrics and solve for the fluid attributes \cite{krori}. Vaidya $\&$ Tikekar \cite{vaidya} proposed a static spherically symmetric model of a superdense star based on an exact solution of Einstein’s equations by prescribing an ansatz (Vaidya–Tikekar ansatz) for the metric functions. It was for the geometry of \lq t = constant' hypersurface and the physical 3-space of the star was spheroidal. Using the Vaidya–Tikekar ansatz several studies have been performed. Gupta $\&$ Kumar \cite{gupta05} observed a particular form of electric field intensity, having positive gradient. He used Vaidya–Tikekar ansatz to generate exact solutions of the field equation in charge analog. Later this form of electric field intensity was used by Sharma et al. \cite{sharma}. Komathiraj $\&$ Maharaj \cite{komaB} additionally accepted a similar articulation to show another kind of Vaidya–Tikekar type star. Bijalwan $\&$ Gupta \cite{bijal11, bijal12} obtained a charged perfect fluid model of Vaidya–Tikekar type stars with more generalized electric intensity. Additionally, some of the other researches on Vaidya–Tikekar stars can be found in literature \cite[see][]{patel97, tikekar98, maha96}. Recently, Kumar et al. \cite{kumar1} used the Vaidya-Tikekar metric potential to explore a class of charged compact objects filled with self-gravitating, charged, isotropic fluids.

Above mentioned literature survey motivate us to perform this research work on the subsequent line of action. In this research paper, our objective is to get an exact solution of the field equations for a static spherically symmetric fluid sphere. The matter distribution is charged with isotropic pressures. This paper is organized as follows: Following a quick introduction in section 1, we have introduced the Einstein-Maxwell field equations for the static charged fluid spheres in general relativity in section 2. In section 3, we have proposed a new model to solve the system of equations analytically. For this, we have used Vaidya $\&$ Tikekar \cite{vaidya} ansatz for the metric potential and acquired the expression for density and pressure. In section 4, we have discussed the requirements for a well-behaved solution. Boundary conditions are discussed in section 5. In section 6, we are going to investigate the applicability of the model and obtained results will be compared with observational data. We have done the stability analysis of the obtained model in this section. Finally, section 7 is dedicated to concluding remarks.
\label{}
\section{Basic field equations in Schwarzschild's canonical coordinates} 
	Let us consider the metric in Schwarzschild coordinates $(x^i) = (t,r,\theta,\phi)$ to describe the interior of static and  spherically symmetric stellar system 
\\\begin{equation}
	ds^2=e^{\nu(r)}dt^2-e^{\lambda(r)}dr^2-r^2(d\theta^2+sin^2\theta d\phi^2),
	\label{metric}
\end{equation}
where the metric potentials $\nu(r)$ and $\lambda(r)$ are arbitrary functions of radial coordinate $r$.  These potentials uniquely determine the surface redshift and gravitational mass function respectively. The signature of the space time taken here is (+,-,-,-).  
The Einstein-Maxwell field equations for obtaining the hydrostatic stellar structure of the charged sphere can be written as 
\begin{equation}
	-\kappa(T_{j}^{i}+E_{j}^{i})=R_{j}^{i}-\frac{1}{2}R\delta_{j}^{i}=G_{j}^{i},
	\label{ife}
\end{equation} 
where $\kappa= \frac{8\pi G}{c^4}$,  $G$ here stands for gravitational constant and $c$ is the speed of light, $R_{j}^{i}$ and $R$ represent Ricci Tensor and Ricci Scalar respectively. Since we are assuming that matter within the star is a charged perfect fluid, the corresponding energy-momentum tensor $T_{j}^{i}$ and electromagnetic field tensor $ E_{j}^{i}$ will be
\begin{equation}
	T_{j}^{i} = (\rho + p) v^{i} v_{j }-p\delta_{j}^{i} 
\end{equation}
and
\begin{equation}
	E_{j}^{i} = \frac{1}{4\pi}(-F^{im}F_{jm}+\frac{1}{4}F^{mn}F_{mn}),
\end{equation}
where, $\rho(r)$ is the energy density, $p(r)$ is the isotropic pressure, $F_{ij}$ is anti-symmetric electromagnetic field strength tensor defined as
$F_{ij} = \frac{\partial A_{j}}{\partial x_{i}}-\frac{\partial A_{i}}{\partial x_{j}}$ which satisfies Maxwells equations, 
\begin{center}
	$F_{ik,j}+F_{kj,i}+F_{ji,k}=0$ and $[\sqrt{-g} F^{ik}]_{,k}=4\pi J^{i}\sqrt{-g}$
\end{center}
Here $A_j = (\phi(r), 0, 0, 0)$ is the potential and $J^{i}$ is the electromagnetic current vector defined as $J^i = \frac {\sigma}{\sqrt{g_{44}}}\frac {dx^i}{dx^4}= \sigma \nu^i$, where $\sigma = e^{(\nu/2)}J^0$ represents the charge density, $g$ is the determinant of the metric $g_{ij}$ which is defined by $g=-e^{\nu+\lambda} r^4 sin^{2}\theta$ and $J^0$ is the only non-vanishing component of the electromagnetic current $J^i$ for the static spherically symmetric stellar system. Since the field is static, we have $\nu=(0,0,0,\frac{1}{\sqrt{g_{44}}})$.

Also, the total charge within a sphere of radius $r$ is given by 
\begin{equation}
	q(r) = r^2E(r) = 4\pi \int_{0}^r J^0 r^2 e^{(\nu+\lambda)/2}dr
	\label{dense}
\end{equation}
where, $E(r)$ is the intensity of the electric field. 

Thus, for the spherically symmetric metric (\ref{metric}) the Einstein field equation (\ref{ife}) provides the following relationship:
\begin{eqnarray}
	\frac{\lambda '}{r}e^{-\lambda}+\frac{1-e^{-\lambda}}{r^2}&=&c^2 \kappa \rho + \frac{q^2}{r^4} 
	\label{fe1}\\
	\frac{\nu '}{r}e^{-\lambda}-\frac{1-e^{-\lambda}}{r^2}&=&\kappa p- \frac{q^2}{r^4}
	\label{fe2}\\
	\Big(\frac{\nu''}{2}-\frac{\lambda' \nu'}{4}+\frac{\nu'^2}{4}+\frac{\nu'-\lambda'}{2r}\Big)e^{-\lambda}&=&\kappa p+ \frac{q^2}{r^4} 
	\label{fe3}
\end{eqnarray}
Here prime denotes differentiation with respect to $r$. By taking radial and transverse stresses equal at each interior point of the stellar configuration and then using Eqs. (\ref{fe2}) and (\ref{fe3}), we can obtain the condition of pressure isotropy as
\begin{equation}
	\Big(\frac{\nu''}{2}-\frac{\lambda' \nu '}{4}+\frac{\nu'^2}{4}-\frac{\nu'+\lambda'}{2r}-\frac{1}{r^2}\Big)e^{-\lambda}+\frac{1}{r^2}= \frac{2q^2}{r^4} 
	\label{fe4}
\end{equation}
We can get the definition of charged density $\sigma$ by substituting this value in eq. (\ref{dense}).

We are using geometrized units, and thus we have taken $G=c=1$ throughout the discussion. 

\section{Proposed Model}
In seeking solutions to Einstein's field equations for isotropic fluid matter, we have five unknown functions of $r$, namely, $q(r)$, $\rho(r)$, $p(r)$, $\nu(r)$ and $\lambda(r)$. Since we have 3 equations (\ref{fe1},\ref{fe2}\& \ref{fe4}) and 5 unknown functions, one has to specify two variables a priori to solve these equations analytically.  Let's consider the widely used Vaidya-Tikekar \cite{vaidya} metric potential

\begin{equation}
	e^\lambda=\frac{K(1+Cr^2)}{K+Cr^2}
	\label{lambda}
\end{equation}
where $C$  and $K$ are two parameters.
Let's introduce a new variable as
\begin{equation}
	e^\nu=Z^2 (r)
	\label{nu}
\end{equation}

The substitution, followed by some computations, leads us to an equivalent form of field equations, which might be helpful to find the exact solutions more efficiently. Using (\ref{fe1}), (\ref{fe2}) and (\ref{fe4}), we have:
\begin{eqnarray}
	c^2\kappa \rho + \frac{q^2}{r^4}&=&\frac{C(K-1)(3+Cr^2)}{K(1+Cr^2)^2}
	\label{fe5}\\
	\kappa p-\frac{q^2}{r^4}&=&\frac{K+Cr^2}{K(1+Cr^2)} \frac{2Z'}{rZ}+\frac{C(1-K)}{K(1+Cr^2)}
	\label{fe6}\\		
	\frac{2q^2}{r^4}&=&\frac{K+Cr^2}{K(1+Cr^2)} \left[\frac{Z''}{Z}-\frac{Z'}{rZ}+\frac{Cr(K-1)}{(K+Cr^2)(1+Cr^2)}\left(Cr-\frac{Z'}{Z}\right)\right]
	\label{fe7}
\end{eqnarray}
Our aim is to consider the charged perfect fluid distribution represented by metric (\ref{metric}) when $0<K<1$.

Let's introduce a new transformation 
\begin{equation}
	X=\sqrt{\frac{K+Cr^2}{1-K}}
	\label{X}
\end{equation}
so as to get a convinient form of the above equations. Substituting the value of $X$ into eq. (\ref{fe7}), we get,
\begin{equation}
	\frac{d^2Z}{dX^2}-\frac{X}{1+X^2}\frac{dZ}{dX}-(1-K)\Big[\frac{1}{1+X^2}+\frac{2K(1+Cr^2)q^2}{C^2r^6} \Big]Z=0
	\label{fe71}
\end{equation}

Let's use the transformation
\begin{equation}
	Z=(1+X^2)^{1/4}Y
	\label{z}
\end{equation}
to convert eq. (\ref{fe71}) into the normal form
\begin{equation}
	\frac{d^2Y}{dX^2}+\psi Y=0,
	\label{nf1}
\end{equation}
where, 
\begin{equation}
	\psi= -\frac{1}{1+X^2}\Big[ 1-K+\frac{2Kq^2(1+Cr^2)^2}{C^2r^6}+\frac{3X^2-2}{4(1+X^2)}\Big]
	\label{nf2}
\end{equation}
 Due to high non-linearity of $\psi$, it is nearly impossible to solve the differential equation (\ref{nf1}) through a direct approach. In order to solve the differential equation (\ref{nf1}), we have considered 
\begin{equation}
	\psi= -\frac{2a}{X^2(a+bX)}
	\label{psi}
\end{equation}
where, $a(\ne0$) and $b$ are real constants. Such a choice of $\psi$, gives a completely new form of electric field intensity. 
Comparision of eqs. (\ref{nf2}) and (\ref{psi}) provides the following definition of electric field  intensity:
\begin{equation}
	E^2=\frac{q^2}{r^4}= \frac{C^2r^2}{2K(1+Cr^2)^2}\Big[ \frac{5}{4}\frac{(1-K)}{(1+Cr^2)}+\frac{2a}{X^2(a+bX)}\frac{1+Cr^2}{(1-K)}+K-\frac{7}{4}\Big]
	\label{charge}
\end{equation}
It will later become evident that, for the star which we have considered here, this electric field intensity is physically viable.
If we put the value of $\psi$ from eq. (\ref{psi}) to eq. (\ref{nf1}), the resulting differential equation will be
\begin{equation}
	X^2(a+bX)\frac{d^2Y}{dX^2}-2aY=0
	\label{nf4}
\end{equation}
Rearranging eq. (\ref{nf4}) in a way such that its comparision with differntial equation $P_{0}\frac{d^2Y}{dX^2}+P_{1}\frac{dY}{dX}+P_{2}Y=R$, leads us to the relationship:  $P_{0}=\frac{a+bX}{X}, P_{1}=0, P_{2}= \frac{-2a}{X^2}$ and $R=0$. 
One can easily verify that this differential equation is exact. Hence, we can write it in the form $ P_{2}-\frac{d P_{1}}{dX}+\frac{d^2P_{0}}{dX^2}=0$. Premitive of this equation is, $P_{0}\frac{dY}{dX}+\Big( P_{1}-\frac{d P_{0}}{dX}\Big)Y=\int RdX +A$. This implies,
\begin{equation}
	\frac{dY}{dX}+\frac{a}{X(a+bX)}Y=A
	\label{halfsol}
\end{equation}
Finally, solution of differential equation (\ref{halfsol}) is given by the expression
\begin{equation}
	Y=\frac{a+bX}{X} \left[ A\frac{a}{b^3}H(X)+B\right]
	\label{soly}
\end{equation}
where, $A$ and $B$ are arbitrary constants and
\begin{equation}
	\mathcal{H}(X)=\frac{\sec^2t-\cos^2t}{2}+\log \cos^2t, \ \ \ \text{with}, \ t= \tan^{-1}\sqrt{\frac{bX}{a}}
\end{equation}

Together eqs. (\ref{z}) and (\ref{soly}) yields,
\begin{equation}
	Z=(1+X^2)^{1/4}\frac{a+bX}{X} \Big [ A\frac{a}{b^3}H(X)+B\Big ]
	\label{solz}
\end{equation}
To achieve the expression for energy density and pressure, let's put eqs. (\ref{charge}) and (\ref{solz}) into eqs. (\ref{fe5}) and (\ref{fe6}), respectively. Hereby, we obtain the following expressions:
\begin{eqnarray}
		c^2\kappa \rho=\frac{C(K-1)(3+Cr^2)}{K(1+Cr^2)^2}- \frac{C^2r^2}{2K(1+Cr^2)^2}\bigg[\frac{5}{4}\frac{(1-K)}{(1+Cr^2)}+\frac{2a}{X^2(a+bX)}\frac{1+Cr^2}{(1-K)}+K-\frac{7}{4}\bigg]
	\label{density}\\
\kappa p=\frac{C^2r^2}{2K(1+Cr^2)^2}\left[\frac{5}{4}\frac{(1-K)}{(1+Cr^2)}+ \frac{2a}{X^2(a+bX)}\frac{1+Cr^2}{(1-K)}+K-\frac{7}{4}\right]+\frac{C(1-K)}{K(1+Cr^2)}+\frac{C(K+Cr^2)}{K(1+Cr^2)}\left[\frac{A_1 A_2+A_3 A_4}{A_2 A_5}\right]
	\label{pressure}
\end{eqnarray}
On differentiating eqs (\ref{density}) and (\ref{pressure}) with respect to $r$, we get graidiant of density and pressure respectively as,
\begin{eqnarray}
	c^2\kappa \frac{d\rho}{dr}&=&C^2r\left[ N_6-N_7+N_8 \right]
	\label{grad.d}\\
	\kappa \frac{dp}{dr}&=&C^2r\left[ \frac{K+Cr^2}{K(1+Cr^2)} \frac{N(r)}{A_2 A_5}+\frac{2(1-K)}{K(1+Cr^2)^2}\left(\frac{A_1 A_2+A_3 A_4}{A_2 A_5}-1\right)+N_7-N_8 \right]
	\label{grad.p}
\end{eqnarray}
\textbf{See Appendix} for $A_i \ (i=1,2,...,5)$, $N$ and $N_j \ (j=1,2,...,8)$. 
\section{Requirements for Physically Consistent Stellar Model}
For a well-behaved nature of the solution, the pre-requisites are:
\begin{enumerate}
	\item The solution should be free from physical and geometrical singularities, i.e., values of central pressure ($p$) and central density ($\rho$) must be finite and positive, and $e^{\lambda}$ and $e^{\nu}$ must have a non-zero positive value. 
	\item The solution should have positive and monotonically decreasing expressions for energy density and pressure with rise of radius $r$. Mathematically, $\rho\ge 0$, $p \ge 0$, $\frac {d\rho}{dr} \le 0$ and $\frac {dp}{dr} \le 0$. At the stellar boundary ($ r = R$) the radial pressure $p$ should vanish, i.e., $p(R)=0$.
	\item The casualty condition should be obeyed, i.e. velocity of sound should be less than that of light throughout the model. Also, it should be decreasing towards the surface. Besides this, at the center, $\Big( \frac{d\rho}{dr}\Big)$and $\Big( \frac{dp}{dr}\Big)$ must be zero and $\Big( \frac{d^2\rho}{dr^2}\Big)$ and $\Big( \frac{d^2p}{dr^2}\Big)$ must have negative value at the center so that the gradient of density and pressure shall be negative within the radius. The condition $\frac{p}{\rho}<\frac{dp}{d\rho}$ should be valid throughout within the sphere.
	\item The Redshift $z$ should be positive, finite, and monotonically decreasing in nature with the increase of $r$. 
	\item The adiabatic constant $\gamma$ should increase from its lowest value $\frac{4}{3}$ at  center to infinity as we move outwards, for a stable model.
	\item The solution must satisfy the Tolman-Oppenheimer-Volkoff (TOV) equation.
	\item The solution is required to fulfill all the energy conditions simultaneously.
	\item The interior metric functions should match smoothly to the exterior Schwarzschild space-time metric at the boundary.
\end{enumerate}

\section{Boundary conditions and determination of arbitrary constants}
The unique exterior metric for a spherically symmetric charged distribution of matter is the  Reissner-N\"{o}rdstro solution. To explore the boundary conditions, we are going to use the principle that, the metric coefficients and their first derivatives in interior solution and  exterior solution are continuous up to and on the boundary. \\
Consider $r = R$ as the outer boundary of the fluid sphere. The field in the exterior region ($r\ge R$) is described by the  Reissner-N\"{o}rdstro metric 
\begin{equation}
	ds^2 = \Big(1-\frac{2M}{r}+\frac{q^2}{r^2}\Big)dt^2 - \Big(1- \frac{2M}{r}+\frac{q^2}{r^2}\Big)^{-1}dr^2-r^2\big(d\theta^2 + sin^2\theta d\phi^2\big),
	\label{extmetric}
\end{equation}
where, $ M= m( R)$, total gravitational mass, is actually a constant, and defined by $M=\xi(R)+\zeta(R)$, with
\begin{align*}
	\zeta(R)=\frac{\kappa}{2}\int_{0}^{R}\rho r^2 dr \ \ and \ \ \xi(R)=\frac{\kappa}{2}\int_{0}^{R}r \sigma q e^{\lambda/2}dr 
\end{align*}	

Here, $\xi(R)$ is the mass and $\zeta(R)$ is the mass equivalence to electromagnetic energy of distribution \cite{florides}.

Applicable boundary conditions  are :
\begin{enumerate}
	\item The interior metric (\ref{metric}) should join smoothly at the surface of spheres (r = R) to the exterior metric (\ref{extmetric}).
	\item Pressure $p(r)$ should vanish at $r = R$.
\end{enumerate}
Arbitrary constants A and B can be obtained using the boundary conditions. The continuity of $e^{\nu}$, $e^{\lambda}$ and $Q$ at $r = R$  implies that, 
\begin{equation}
	e^{\nu(R)}=Z^2(R)=1- \frac{2M}{R}+\frac{Q^2}{R^2}, \ \ e^{-\lambda(R)} =1- \frac{2M}{R}+\frac{Q^2}{R^2},\ \  Q=q(R), \  \ p(R)=0
	\label{bcon}
\end{equation}
Using $P(R)=0$, we can easily obtain $\frac{B}{A}$. Then, along with condition $e^{\nu(R)}=e^{-\lambda(R)}$ it will give us $A$ and $B$ as,
\begin{eqnarray}
	\nonumber	A&=&-\frac{1}{\sqrt{K}(1+{X_1}^2)^{5/4}}\Big [\frac{A_{11} X_1^2+(J+1) A_{51}}{(a+bX_1) A_{31} A_{41}}\Big]
	\label{A}\\
	\nonumber	B&=&\frac{1}{\sqrt{K}(1+{X_1}^2)^{5/4}}\Big[\frac{X_1^2}{a+bX_1}+\frac{a}{b^3}\frac{A_{11} X_1^2+(J+1) A_{51}}{(a+bX_1) A_{31} A_{41}}H(X_1)\Big]
	\label{B}
\end{eqnarray}
\textbf{See Appendix} for values of $A_{i1} \ (i=1,3,4,5)$, $J$ and $X_1$.

	\section{Compatibility of the model with the Realistic Compact Star}
The solutions discussed in this paper can be used to model a relativistic star. In this Section, we have critically verified our model by performing mathematical analysis and plotting several graphs. This  indicates that the result overcomes all the barriers of physical tests.\\

The expression for the energy density in eq. (\ref{density}) implies that at  $r = 0$,
\begin{equation}
	\rho_0=\frac{3C(K-1)}{8\pi K}
	\label{centdens}
	\end {equation}
	As, $0< K < 1$, the central density $\rho_0$  will be positive iff  we take $C<0$. 
	
	It is also necessary to restrict $C$ so that the transformation (\ref{X})  remains physically acceptable throughout the configuration. For this, we require $|C|\le\frac{K}{R^2}$. However, the function we have chosen for $\psi$ in eq. (\ref{psi}) requires that $C\ne-\frac{K}{R^2}$. Thus, in our model, we have to consider $|C|<\frac{K}{R^2}$. Also, we can say that $K$ and $R$ characterize the geometry of the star.\\
	
	The physical acceptability of the model has been examined by plugging the mass and radius of observed pulsar as input parameters. To validate our model, we have considered the pulsar PSR B0943+10, a low-mass bare quark star of radius $r\sim2.6 \  km$ and mass $M\sim0.02 {M}_\odot$ \cite{psr}. Using these values of mass and radius as an input parameter, the required physical conditions have been utilized to determine the constants as $C=-4.01893\times10^{-5} km^{-2}$, $K=0.0119$, $a=0.001$ and $b=0.029$. Furthermore, using boundary conditions values of arbitrary constants used in eq. (\ref{solz}) are obtained as $ A=- 0.177729  $ and $B= 29.657254$. 
		\begin{table}[h!]
		\centering
		\caption{Structural properties of  PSR \textquotedblleft B0943+10\textquotedblright within its radius.}
		\begin{tabular}{|cccccccc|}
			\hline
			$r/R$ & $q(km)$ &$\rho (km^{-2})$& $p (km^{-2})$& $p/\rho$ & $dp/c^2d\rho$&$z$ &$\gamma$\\
			\hline\rule[-1ex]{0pt}{3.5ex}
			$0.0$ & $0.0$ &$3.98537\times 10^{-4}$&$4.42866\times10^{-8}$&$ 0.000111 $& $0.449988$&$0.017248$&$4047.839101$\\
			\hline\rule[-1ex]{0pt}{3.5ex}
			$0.2$ & $0.0002314$ &$3.98513\times10^{-4}$&$4.13532\times10^{-8}$&$ 0.000104 $& $0.445668$&$0.017019$&$4293.081396$\\
			\hline\rule[-1ex]{0pt}{3.5ex}
			$0.4$ & $0.0018512$ &$3.98448\times10^{-4}$&$3.31219\times10^{-8}$& $0.000083$ & $0.432779$&$0.016329$& $5203.995995$\\
			\hline\rule[-1ex]{0pt}{3.5ex}
			$0.6$ &$0.006266$&$3.98337\times10^{-4}$&$2.13164\times10^{-8}$& $0.000052$& $0.411541$&$0.015178$& $7686.893767$\\
			\hline\rule[-1ex]{0pt}{3.5ex}
			$0.8$ &$0.0149214$&$3.98178\times10^{-4}$&$8.86689\times10^{-9}$&$0.000022$ & $0.382306$&$0.013562$& $17159.507244$\\
			\hline\rule[-1ex]{0pt}{3.5ex}
			$1.0$ & $0.029315$ &$3.97966\times10^{-4}$& $0.0$& $0.0$& $0.345547$& $0.011477$&$Inf.$\\
			\hline
		\end{tabular}
		\label{t1}
	\end{table}															
	\begin{table}[h!]
		\centering
		\caption{Obtained numerical values for charge at surface, central density, surface density, central pressure and mass-radius ratio of compact star  PSR \textquotedblleft B0943+10\textquotedblright.}
		\begin{tabular}{|ccccc|}
			\hline
			$q_s$(C) & $\rho_0(g/cm^3)$ &$\rho_s (g/cm^3)$& $p_0$(Pa)& $M/R$ \\
			\hline\rule[-1ex]{0pt}{3.5ex}
			$3.41784\times10^{18}$ & $5.37433\times10^{14}$ &$5.36663\times10^{14}$& $5.3749\times10^{30}$& $0.011346 $\\
			\hline
		\end{tabular}
		\label{t2}
	\end{table}
	
	Here we have used, $G=6.674\times10^{-11} \ N/ms^2$, $c=3\times10^8 \ m/s$, $1\ M_\odot =1.475 \ km$, to obtain numerical values of physical quantities and have multiplied charge by $1.1659\times10^{20}$ to convert it from relativistic unit ($km$) to  coulomb.
	
	\subsection{Physical Analysis of the model}

	\subsubsection{ Regularity and Reality Conditions}
	\begin{figure}[H]
		\begin{center}
			\begin{minipage}{0.33\textwidth}
			\includegraphics[width=5cm]{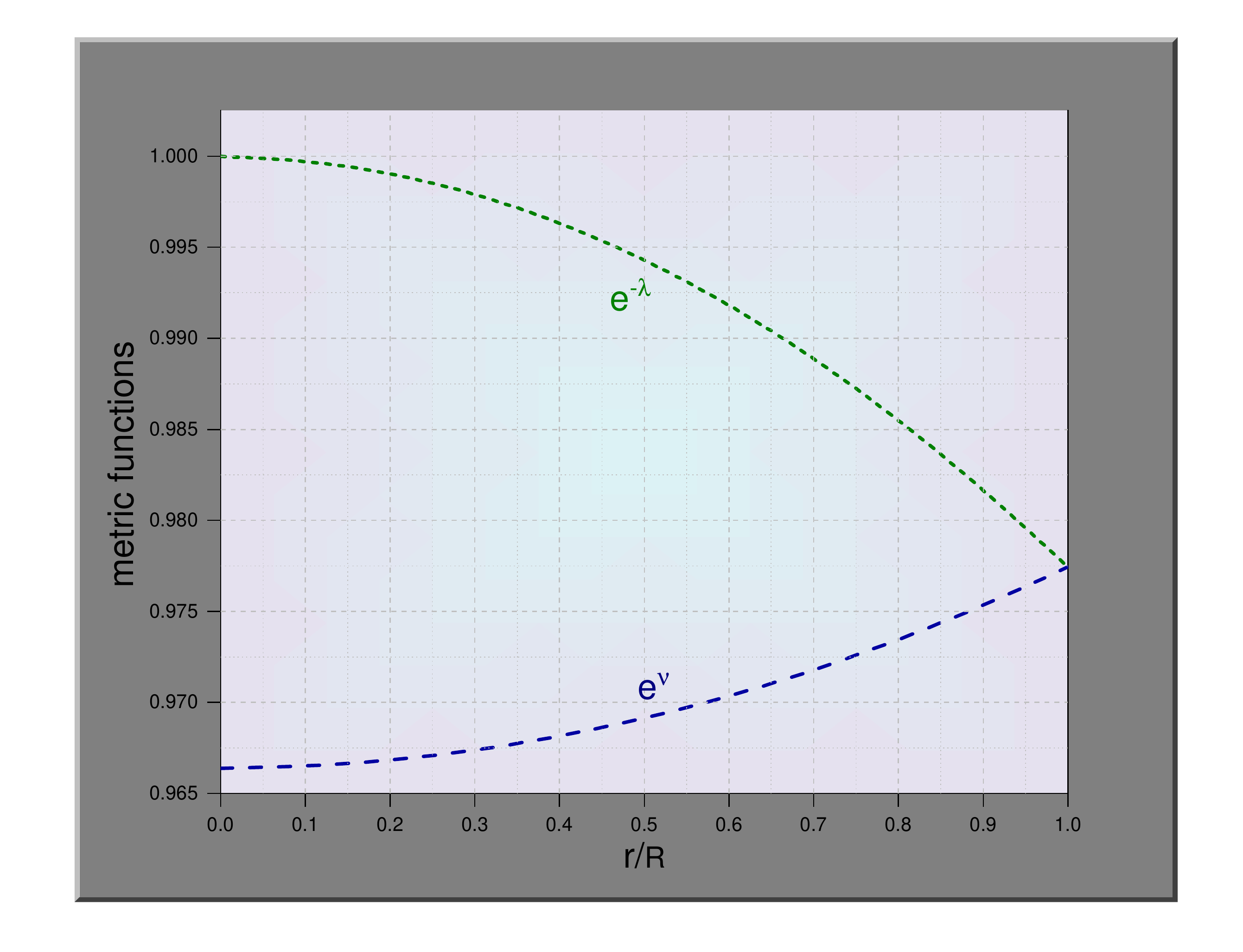} 
			\caption{Variation of $e^{-\lambda}$(green) and $e^{\nu}$(blue) with respect to fractional radius (r/R)}\label{lambnu}
		\end{minipage}
				\begin{minipage}{0.33\textwidth}
			\includegraphics[width=5cm]{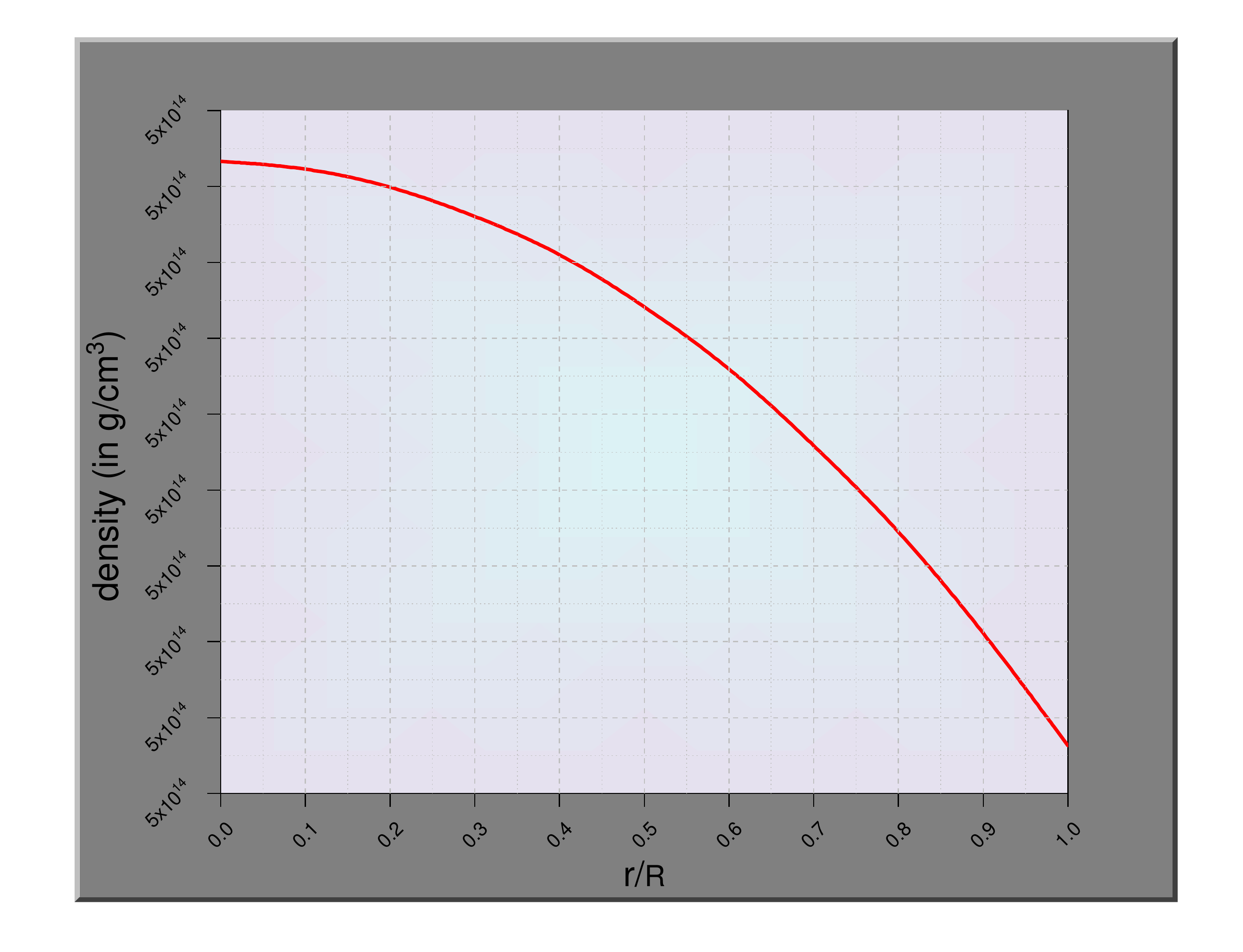} 
			\caption{Variation of density with respect to fractional radius (r/R).}\label{dens}
		\end{minipage}
		\begin{minipage}{0.33\textwidth}
			\includegraphics[width=5cm]{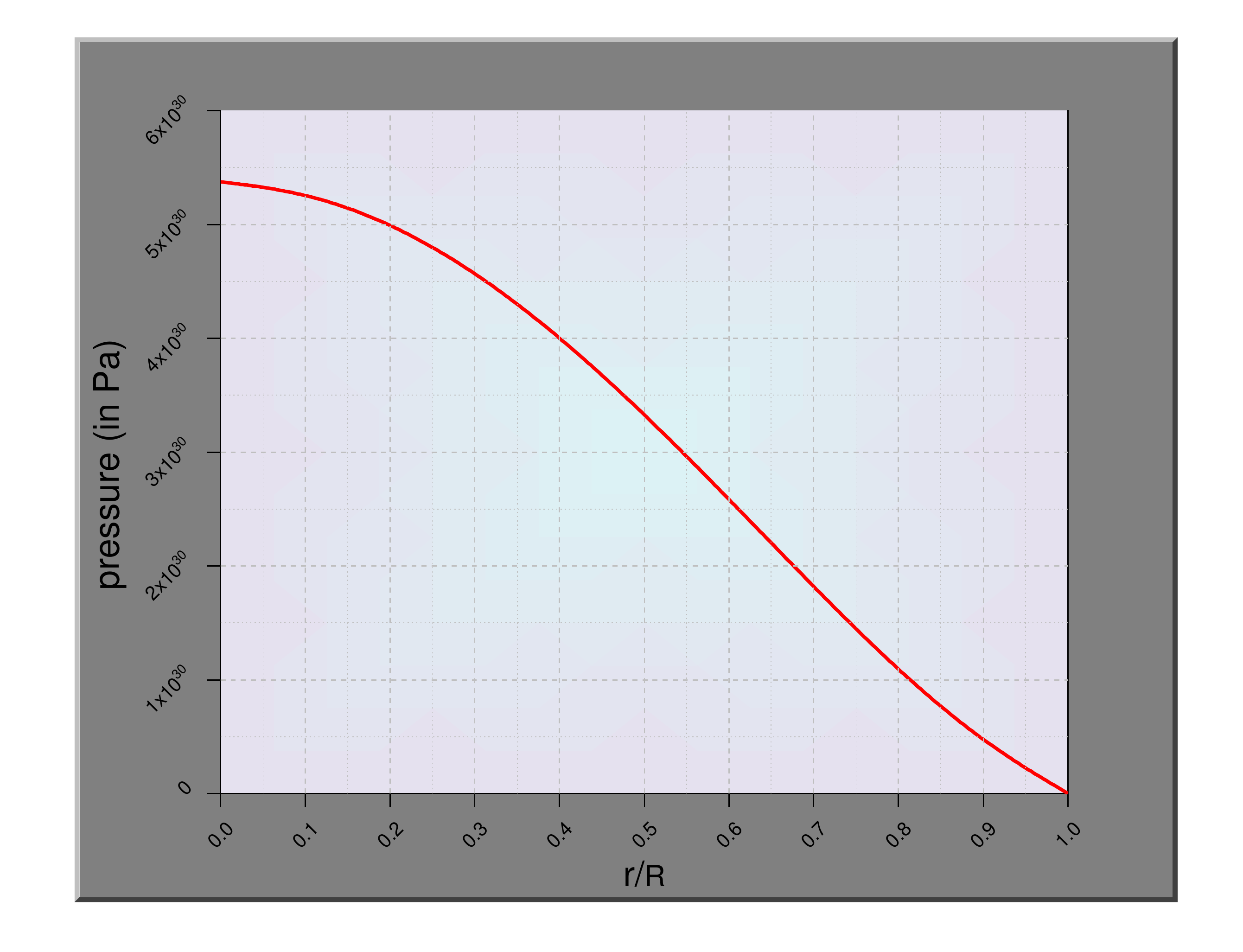} 
			\caption{Variation of pressure with respect to fractional radius (r/R).}\label{press}
		\end{minipage}
	\end{center}
	\end{figure}
	
	\begin{enumerate}
		\item[(i)]The profiles of $e^{-\lambda}$ and $e^\nu$ for PSR B0943+10 in Fig. (\ref{lambnu}) show that the metric potentials are free from physical and geometrical singularities, regular at the centre (i.e., the metric potentials are positive and finite at the centre). $e^{\lambda}$ and $e^\nu$ are monotonically increasing with the radius inside the star. Also, both $e^{-\lambda}$ and $e^\nu$ coincides at the surface. In short, the the behavior of metric functions is consistent with the requirements for a physically acceptable model. 									
		\item[(ii)]
		The graph in Fig. (\ref{dens})  indicates that the energy density is positive with a maximum value at the centre and decreasing in nature throughout the star. Also,  We can see in Fig. (\ref{press})  that  the pressure is  monotonically decreasing towards the surface. At the centre, it is finite and vanishes at the boundary of the star.
	\end{enumerate}					
	\begin{figure}[h!]
		\centering
			\includegraphics[width=5cm]{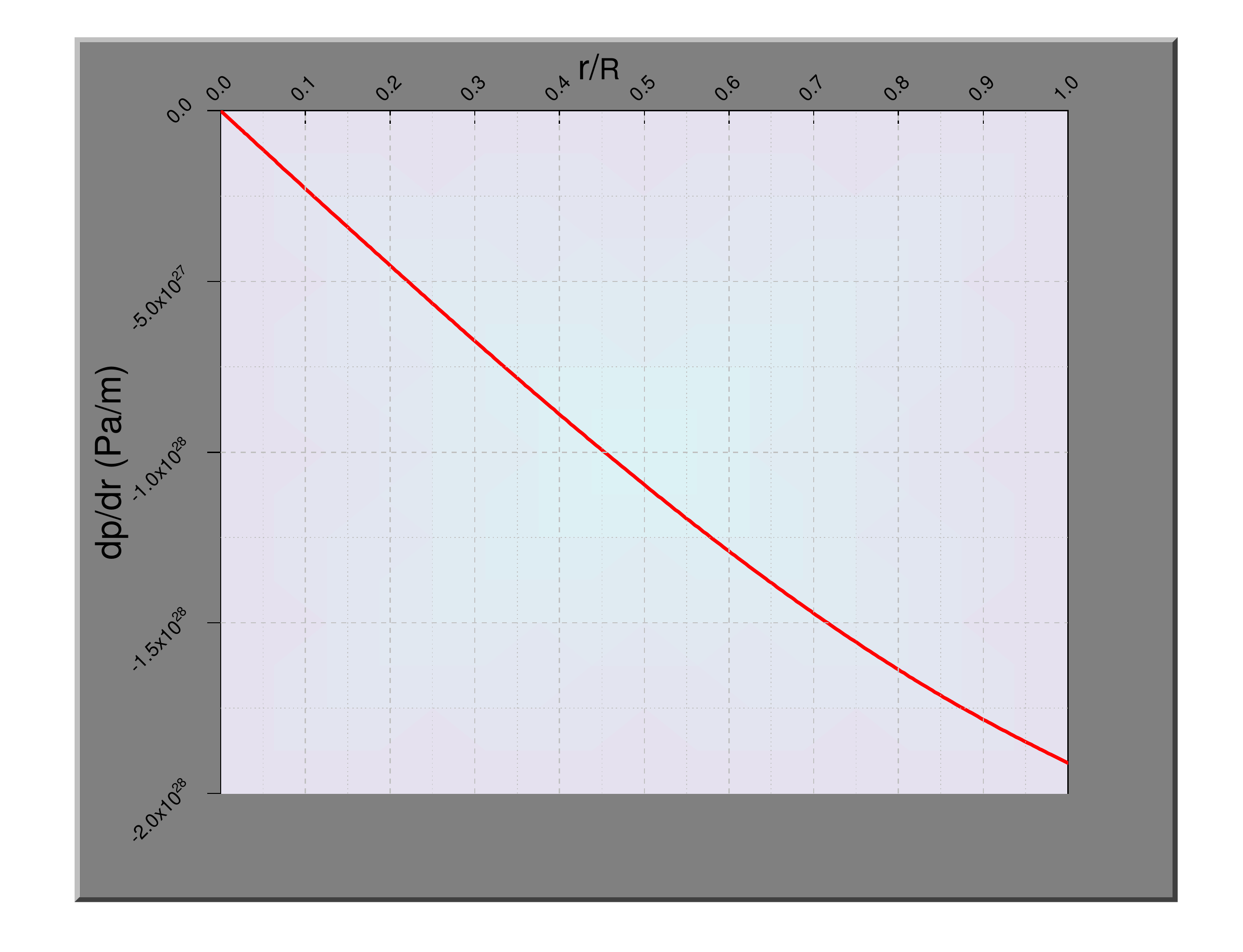} \includegraphics[width=5cm]{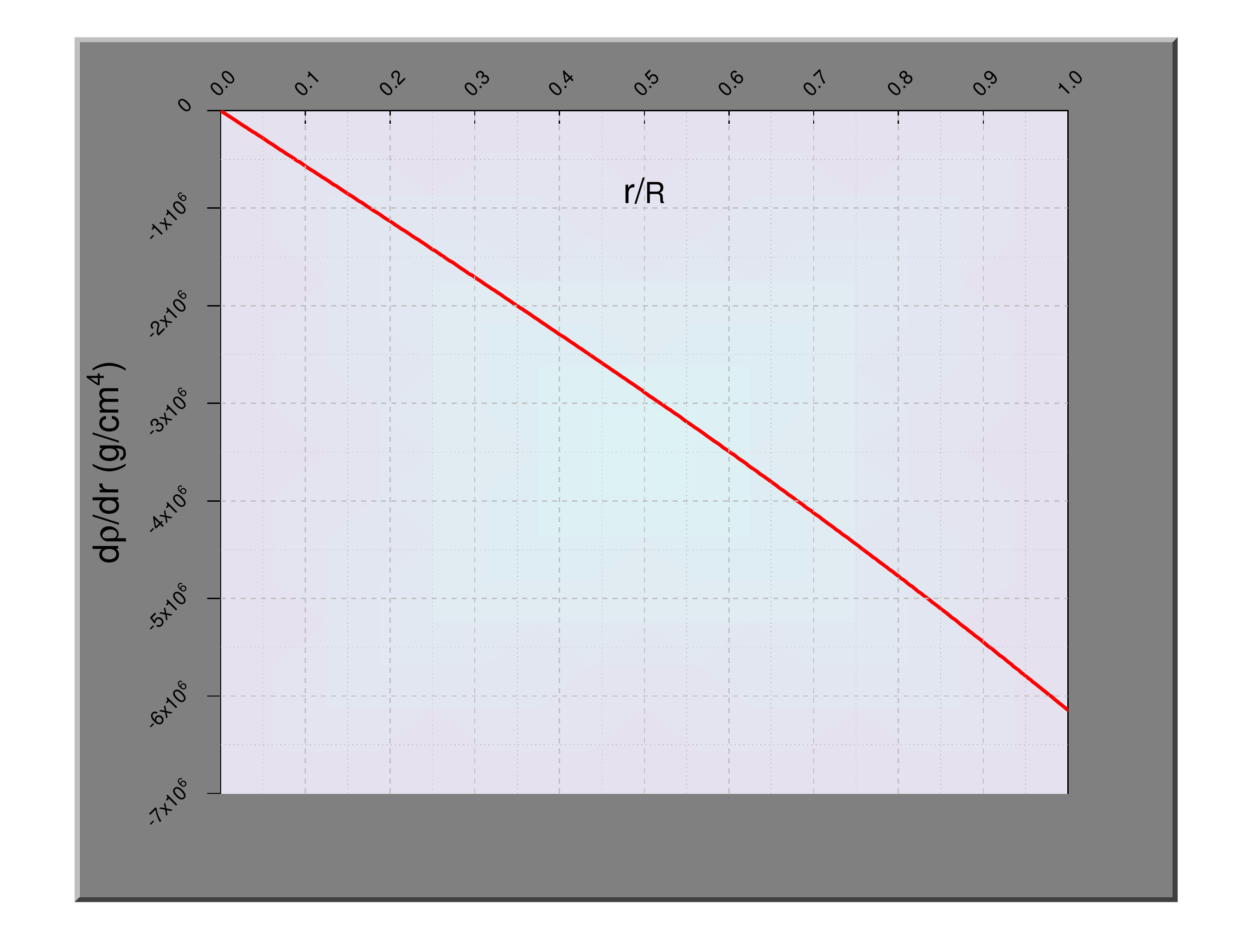}
			\caption{Variation of pressure gradient (left) and density gradient (right) with respect to fractional radius (r/R).}\label{nd}
	\end{figure}
	
	\subsubsection{Causality condition}
		\begin{figure}[h]
		\centering
		\begin{minipage}{0.33\textwidth}
			\includegraphics[width=5cm]{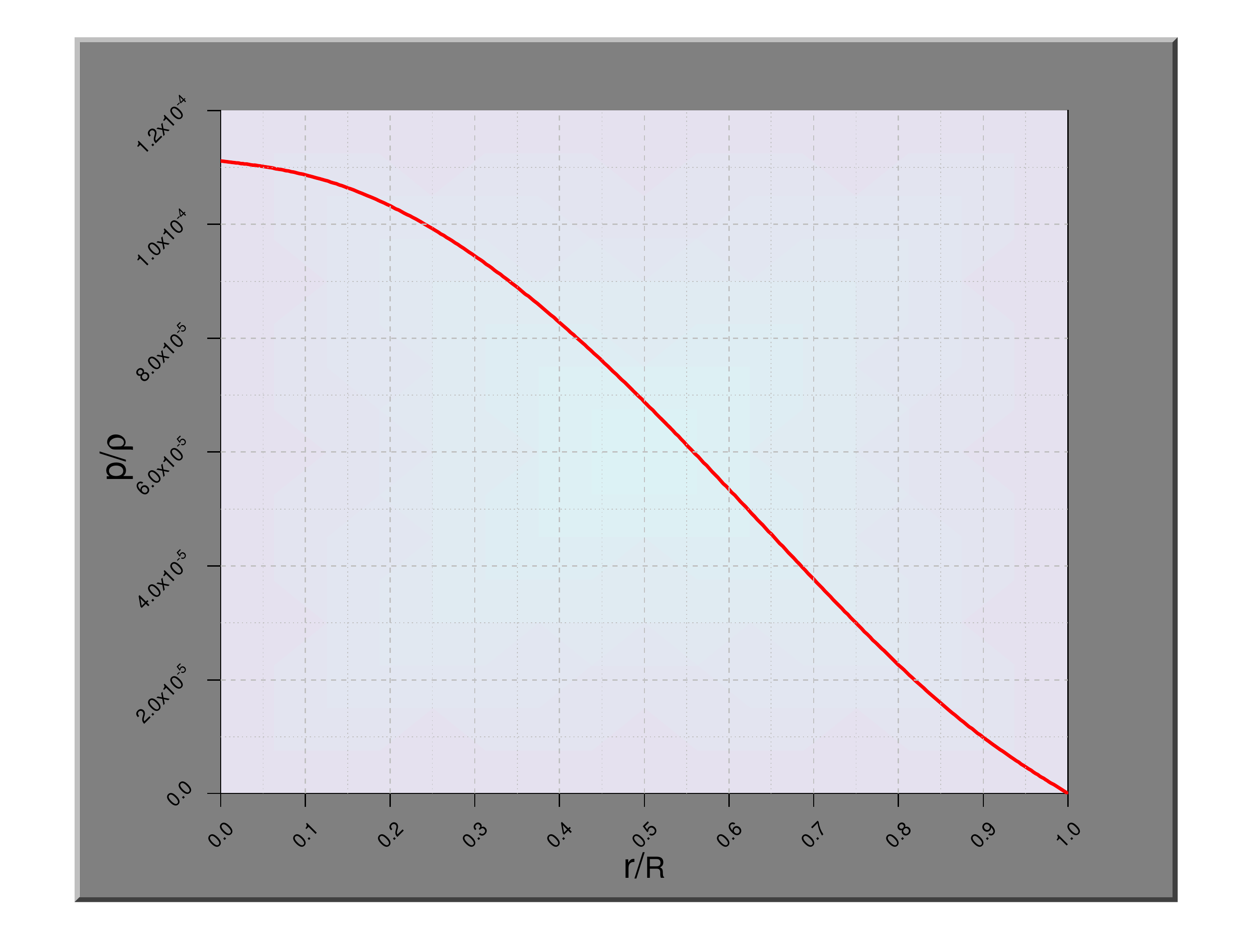}
			\caption{Variation of ratio of pressure to that of density with increase in radius (r/R).}\label{ratio}
		\end{minipage}
		\begin{minipage}{0.47\textwidth}
			\includegraphics[width=5cm]{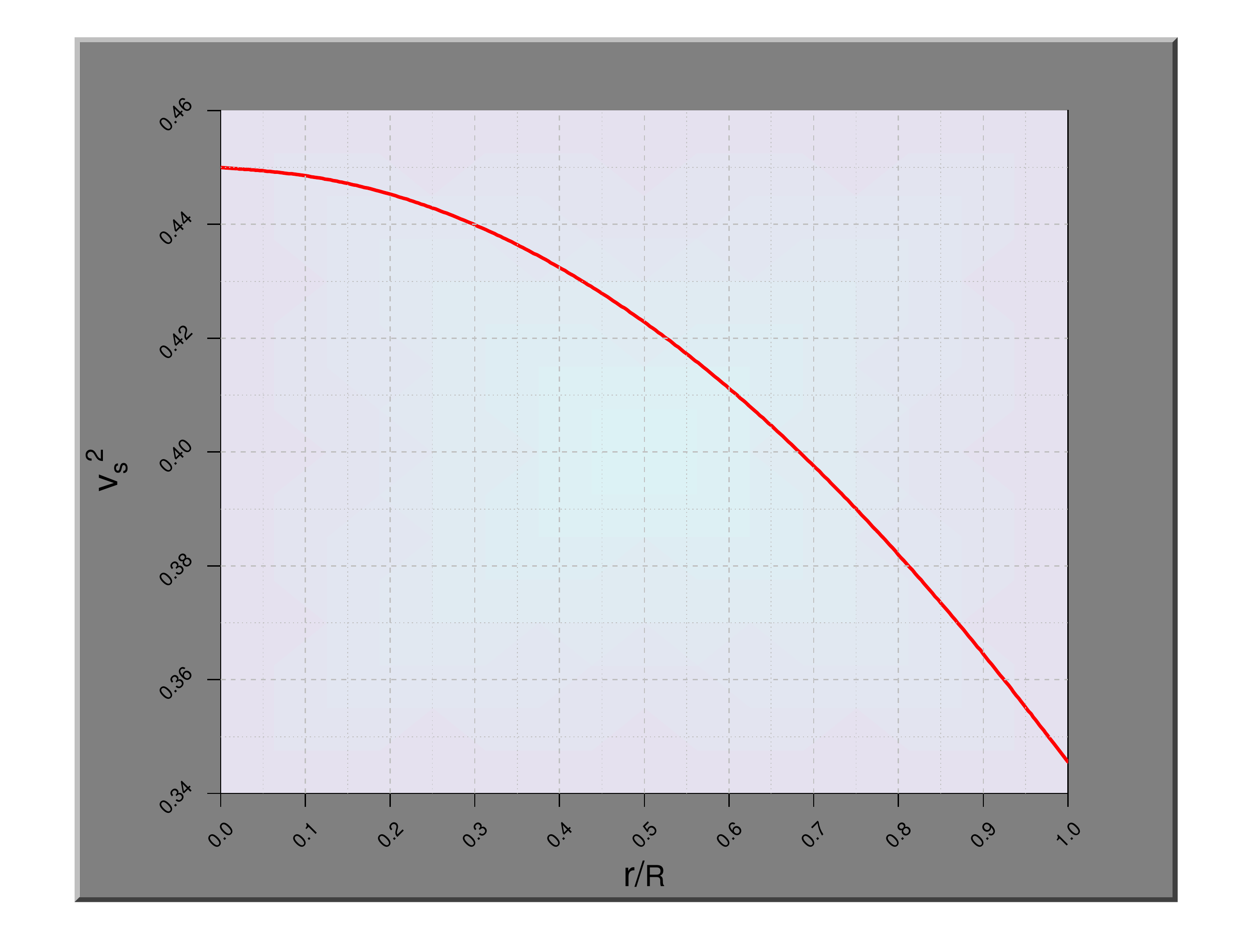} 
			\caption{Variation of square of velocity of sound with respect to fractional radius (r/R).}\label{v}
		\end{minipage}
	\end{figure}
	For a physically acceptable isotropic model,  the square in sound speed ${v_s}$  must  be less than 1 in the star's interior, i.e., $0\le {v_s}^2=\frac{dp}{d\rho}\le1$ \cite{herrera}. This condition is known as the causality condition.  Fig. (\ref{v}) shows that, for our charged isotropic model, the velocity of sound remains less than the speed of light inside the star and decreases with an increase of $r$. 
	
	It is very clear from fig (\ref{ratio}) and fig (\ref{v}) that the ratio $\frac{p}{\rho}$ is less than $\frac{dp}{d\rho}$ throughout the stellar model. One can also verify this through Table \ref{t1}. As we can see in fig (\ref{nd}), gradient of pressure and density are zero at centre and have negative values at every other point in the region.
	
	\subsubsection{ Energy Conditions}
\begin{figure}[H]
		\begin{center}
			\begin{minipage}{0.47\textwidth}
				\includegraphics[width=5cm]{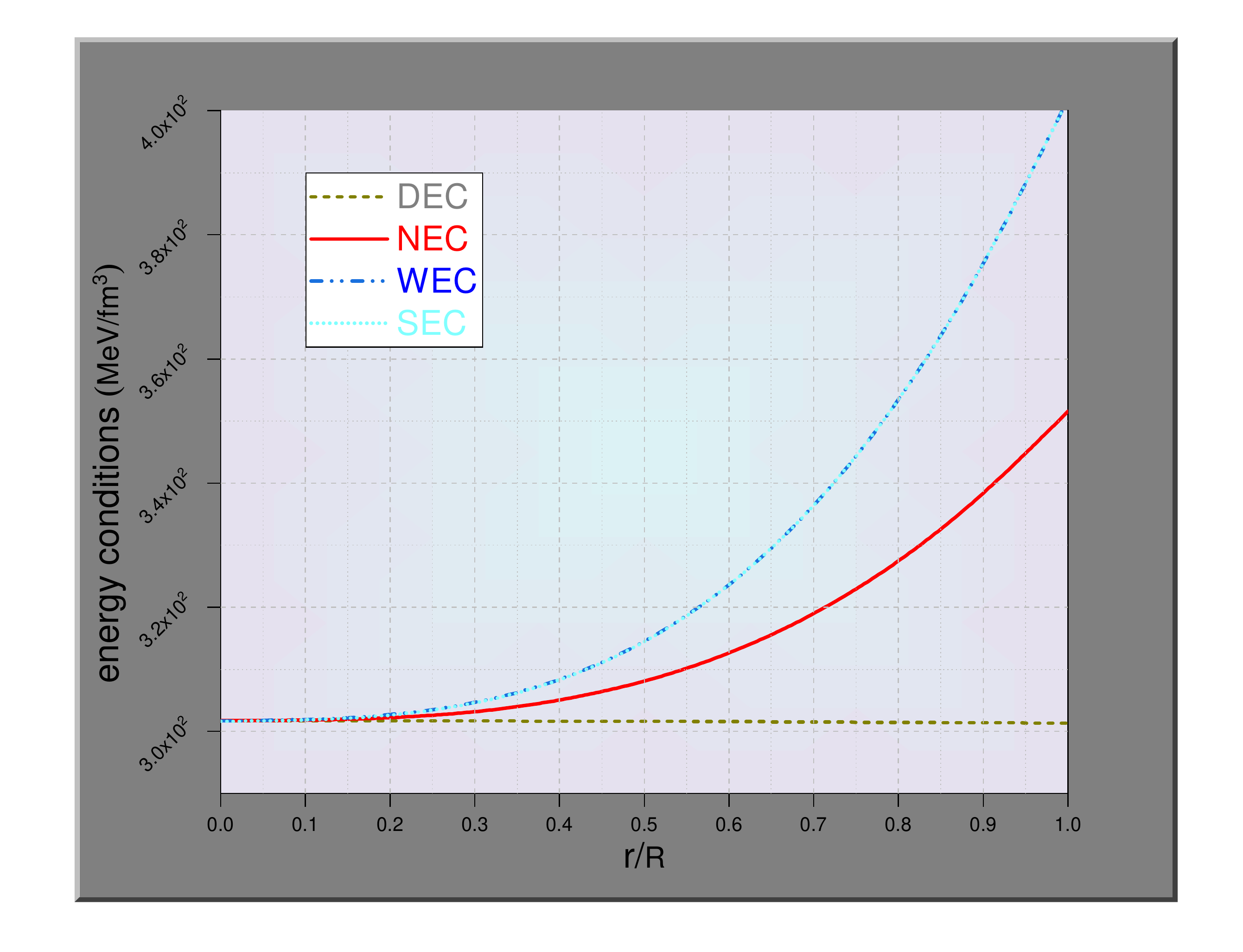} 
				\caption{Energy Conditions within stellar configuration of PSR B0943+10.}\label{ec}
			\end{minipage}
			\begin{minipage}{0.47\textwidth}
				\includegraphics[width=5cm]{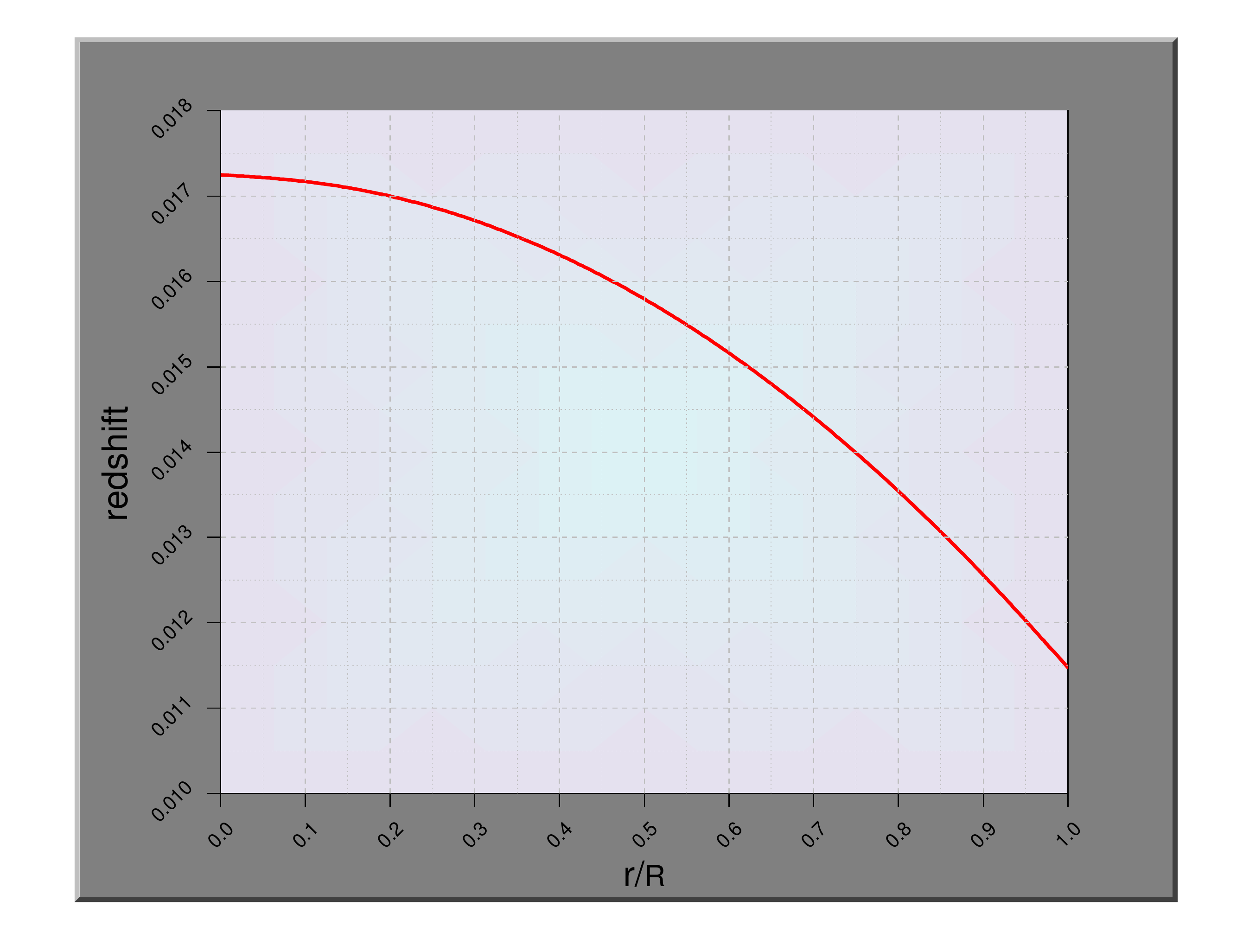} 
				\caption{Variation of redshift with respect to fractional radius (r/R).}\label{red}
			\end{minipage}
		\end{center}
	\end{figure}

	In the study of stellar configurations describing charged isotropic matter distributions, it  is necessary to check whether the energy-momentum tensor is well behaved, i.e. positive defined everywhere within the star. For this, the fulfillment of below energy conditions are required \cite{EngCond}: 
	\begin{enumerate}
		\item Dominant energy condition  (DEC): $c^2\rho-p\ge 0$
		\item Null energy condition (NEC): $c^2\rho+\frac{q^2}{\kappa r^4}\ge0$ 
		\item Weak energy condition (WEC): $c^2\rho-p+\frac{2q^2}{\kappa r^4}\ge0$
		\item Strong energy condition (SEC): $c^2\rho-3p+\frac{2q^2}{\kappa r^4}\ge0$
	\end{enumerate}
	The behaviour of these energy conditions for our particular model PSR B0943+10  is shown in Fig. (\ref{ec}). This figure clearly indicates that all the energy conditions in our model are satisfied throughout the interior region of the spherical distribution.
	\subsubsection{Redshift}								
	The gravitational redshift $z$ within the static line element is given by
	\begin{equation}
		z=|e^{\nu(r)}|^{-1/2}-1 =\frac{1}{|Z(r)|} -1
	\end{equation}													
	Note that surface redshift also stablishes the following relationship
	\begin{equation}
		z_s=|e^{\nu(R)}|^{-1/2}-1 = \Big(1-\frac{2M}{R}+\frac{Q^2}{R^2}\Big)^{-1/2}-1
	\end{equation}	
	
	We have central redshift $z_0=0.017248$ and surface redshift $z_s=0.011477$ and as can be seen in Fig.(\ref{red}), the redshift function gradually decreases inside the star. Clearly we have $z_s<2$, which is required for spherically symmetric isotropic fluid spheres  as suggested in \cite{dahl,trauma,bohmer}.
	
	\subsection{ Stability analysis of the model}
	Now, we will examine the stability of our isotropic, charged fluid configuration in the Einstein-Maxwell space-time in the following subsections.
	\subsubsection{ Electric charge in the sphere}
	Any fluid sphere with net charge, contains fluid elements with unbounded proper charged density located at the fluid-vaccum interface. This net charge can be huge ($10^{19}$C) \cite{varela}. Ray et al. \cite{ray} have analyzed the impact of charge in compact stars by considering the limit of the very extreme measure of the charges. They have demonstrated that the global balance of the forces allows a huge charge ($10^{20}$C) to be available in a compact star. 
	
\begin{figure}[h]
		\centering
		\begin{minipage}{0.47\textwidth}
			\includegraphics[width=5cm]{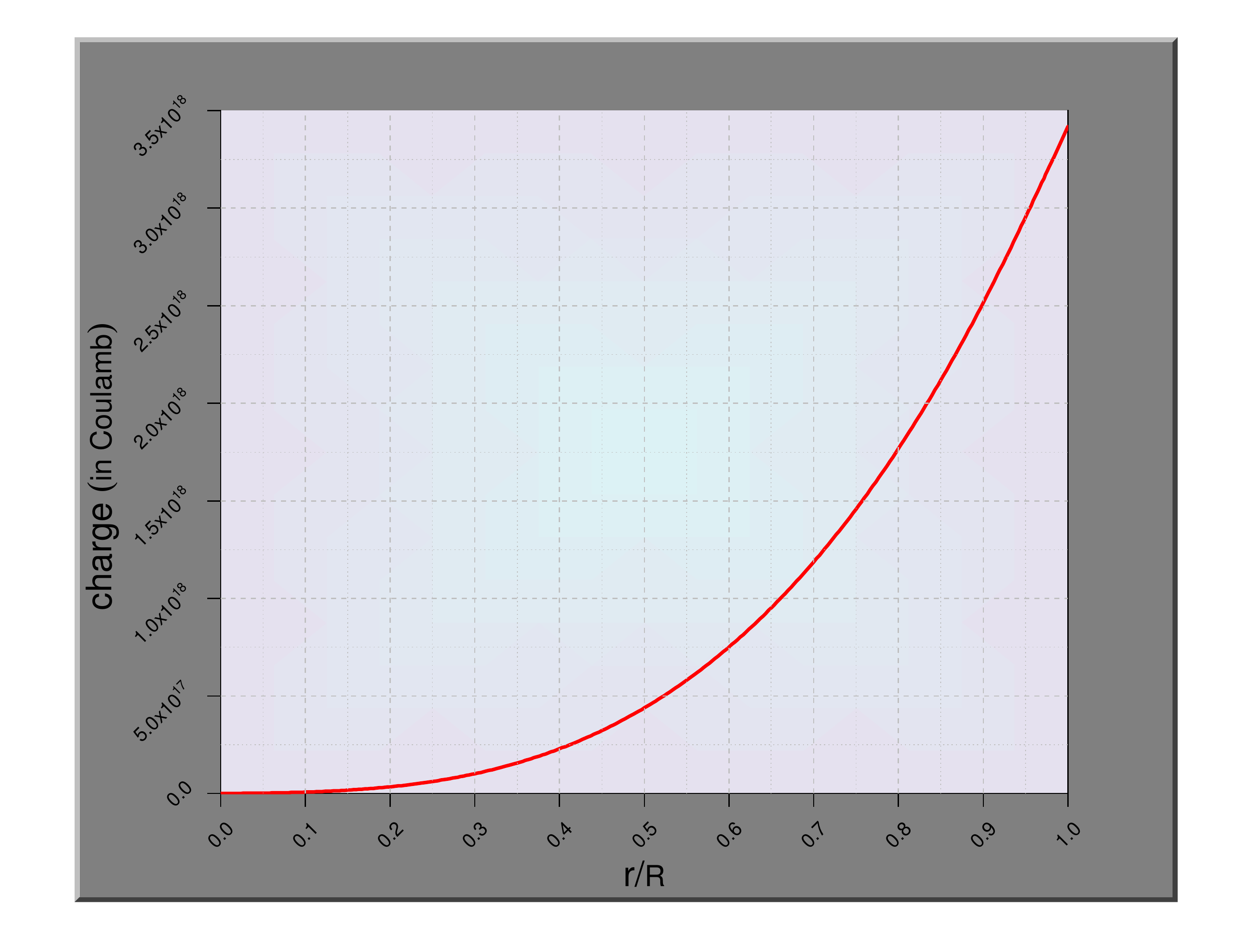} 
			\caption{Behavior of charge within the configuration of PSR B0946+10.}\label{fig:fq}
		\end{minipage}
		\begin{minipage}{0.47\textwidth}
			\includegraphics[width=5cm]{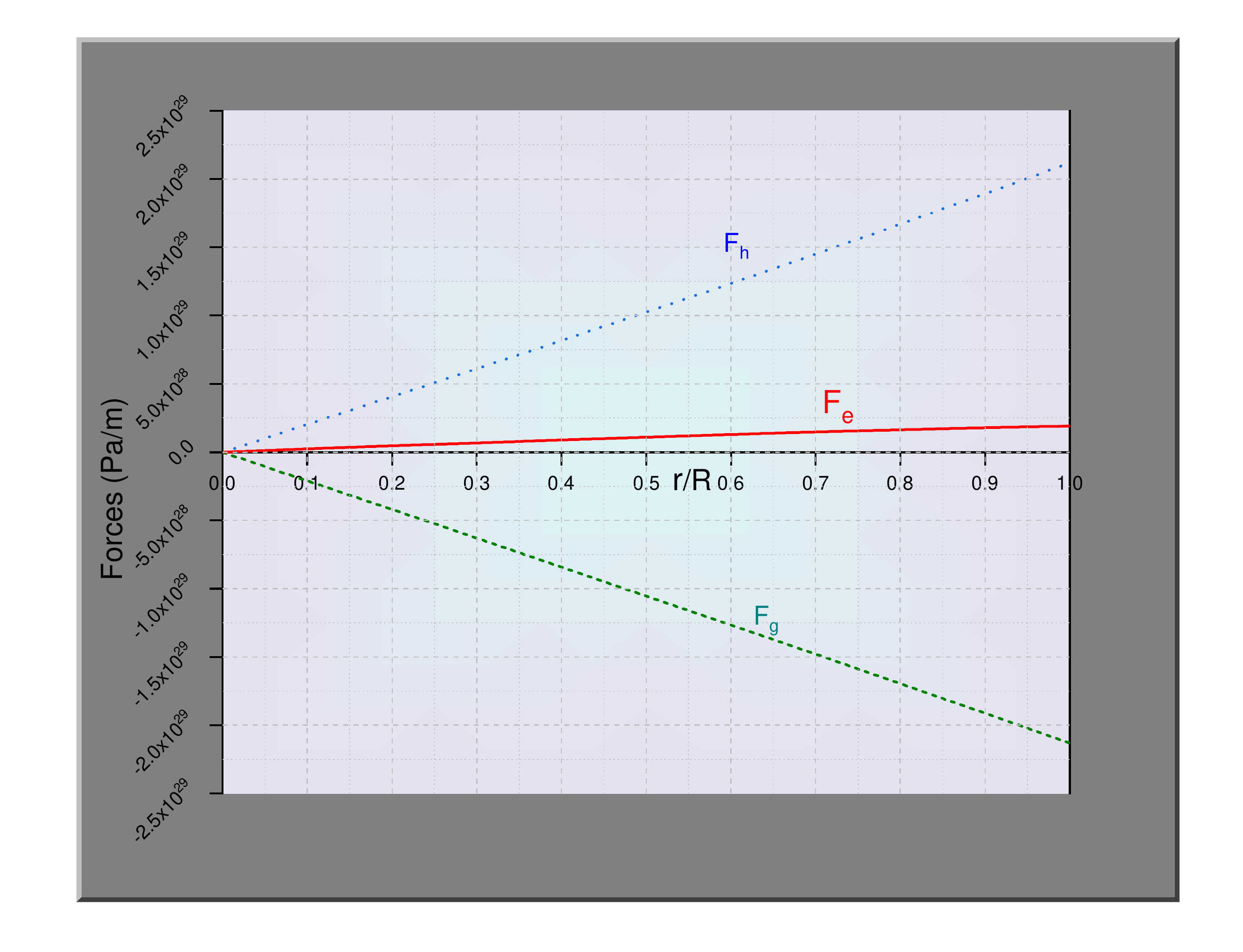}
			\caption{Variations of gravitational force ($F_{g}$),  hydrostatic force ($F_{h}$)  and electric force ($F_{e}$) acting on the system with respect to fractional radius ($r/R$).} \label{fghe}
		\end{minipage}
	\end{figure}

	In Fig. (\ref{fig:fq}), we can observe that the electric field given by eq. (\ref{charge}) is positive and an increasing function with increasing radius. The charge starts from zero at the centre and acquires the maximum value at the boundary.
	In this model, the charge on the boundary is $3.41784\times10^{18}$C. Thus, we can say that, in this model the net charge is effective to balance the mechanism of the force.
	\subsubsection{Equilibrium analysis through TOV equation}
	A star remains in static equilibrium under the forces, namely, gravitational force ($F_{g}$), hydrostatic force ($F_{h}$) and electric force ($F_{e}$). This condition is formulated mathematically as the TOV equation by  Tolman, Oppenheimer $\&$ Volkof \cite{tov1,tov2}.  In the presence of a  charge, the same takes the following form \cite{tov3}
	\begin{equation}
		\frac{-M_{G}(\rho+p)}{r^2}e^{(\lambda-\nu)/2}-\frac{dp}{dr}+\sigma \frac{q}{r^2}e^{\lambda/2}=0 
		\label{tov}
	\end{equation}
	where $M_{G}(r)$ is  the gravitational mass of the star within radius $r$ and is defined by 
	\begin{equation}
		M_{G}(r)=\frac{1}{2}r^2 \nu ' e^{(\nu-\lambda)/2}
	\end{equation}
	Substituting the value of $M_{G}(r)$ in eq. (\ref{tov}), we obtain,
	\begin{equation}
		-\frac{\nu'}{2}(\rho+p)-\frac{dp}{dr}+\sigma \frac{q}{r^2}e^{\lambda/2}=0 
	\end{equation}
	which is equivalent to
	\begin{equation}
		F_{g} + F_{h}+F_{e}=0 
	\end{equation}
	where,
	\begin{eqnarray}
		\nonumber	F_{g}&=&-\frac{\nu'}{2}(\rho+p)= -\frac{Z'}{Z}(\rho+p) = -\frac{C^2r}{16 \pi} \left[\frac{A_1 A_2+A_3 A_4}{A_2 A_5}\right] \left[ \frac{2(K-1)}{K(1+Cr^2)^2}+\frac{K+Cr^2}{K(1+Cr^2)} \frac{A_1 A_2+A_3 A_4}{A_2 A_5}\right] \\
		\nonumber	F_{h}&=&-\frac{dp}{dr}= -\frac{C^2r}{8\pi}\left[ \frac{K+Cr^2}{K(1+Cr^2)} \frac{N(r)}{A_2 A_5}+\frac{2(1-K)}{K(1+Cr^2)^2}\left(\frac{A_1 A_2+A_3 A_4}{A_2 A_5}-1\right)+N_7-N_8 \right] \ \ \& \\
		\nonumber	F_{e}&=&\sigma \frac{q}{r^2}e^{\lambda/2}= \frac{1}{8\pi r^4}\frac{dq^2}{dr}= \frac{C^2r}{8\pi}\left[\frac{3+Cr^2}{K(1+Cr^2)^3}\Big\{ \frac{5}{4}\frac{(1-K)}{(1+Cr^2)}+\frac{2a}{X^2(a+bX)}\frac{1+Cr^2}{(1-K)}+K-\frac{7}{4}\Big\}-N_8 \right].
	\end{eqnarray}						
	
		The graph in fig. (\ref{fghe}) shows that $F_{h}$ and $F_{e}$ are positive and are nullified by $F_{g}$, which is negative to keep the system in static equilibrium.
	\subsubsection{Relativistic Adiabatic Index}
	The adiabatic index, defined as 
	\begin{equation}
		\gamma=\Big(\frac{c^2\rho+p}{p} \Big)\Big(\frac{dp}{c^2 d\rho}\Big)
	\end{equation}
	is related to the stability of an isotropic stellar configuration.

\begin{figure}[h]
		\begin{center}
			\begin{minipage}{0.47\textwidth}
				\includegraphics[width=5cm]{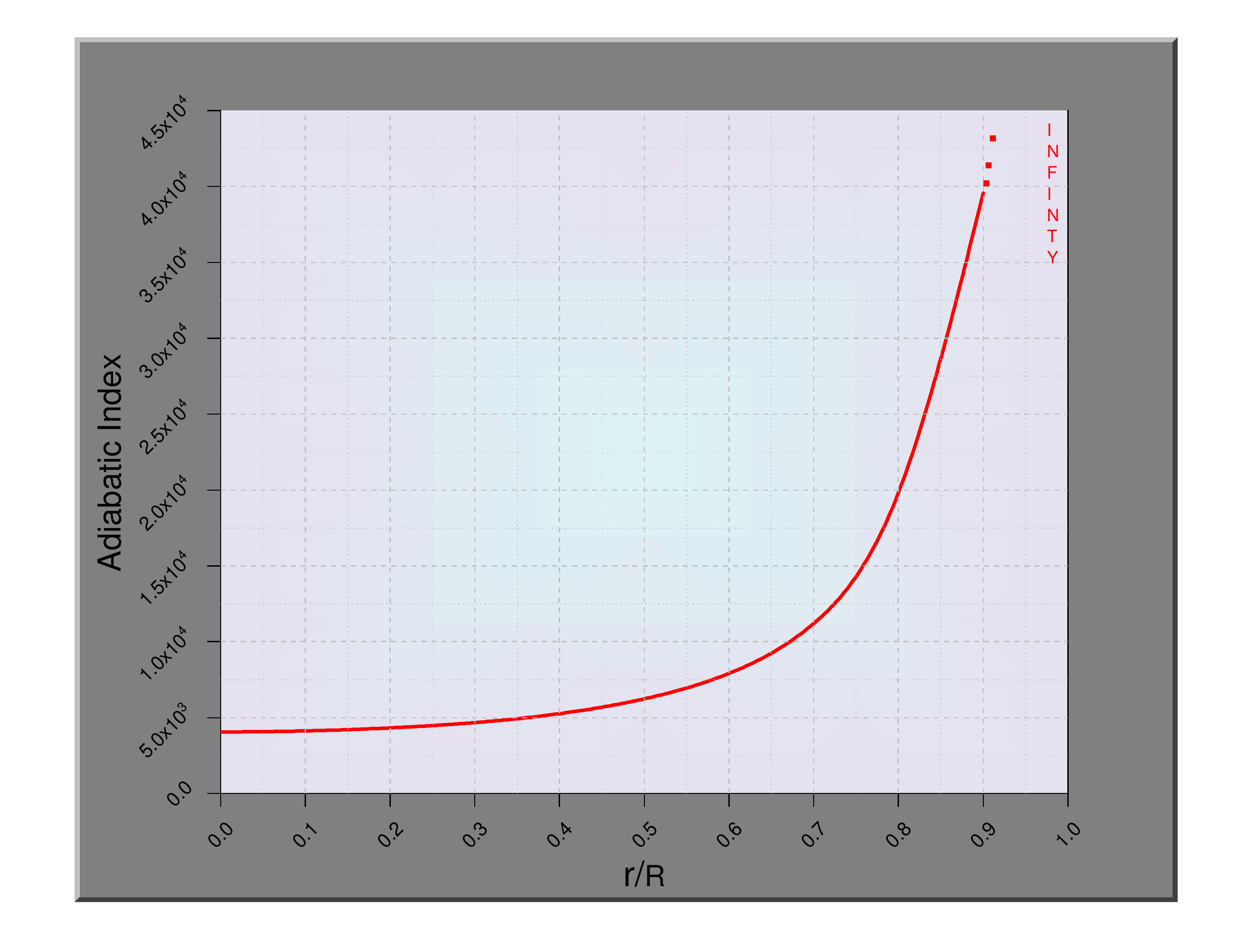} 
				\caption{Nature of adiabatic index with increase in radius.}\label{adia}
			\end{minipage}
			\begin{minipage}{0.47\textwidth}
				\includegraphics[width=5cm]{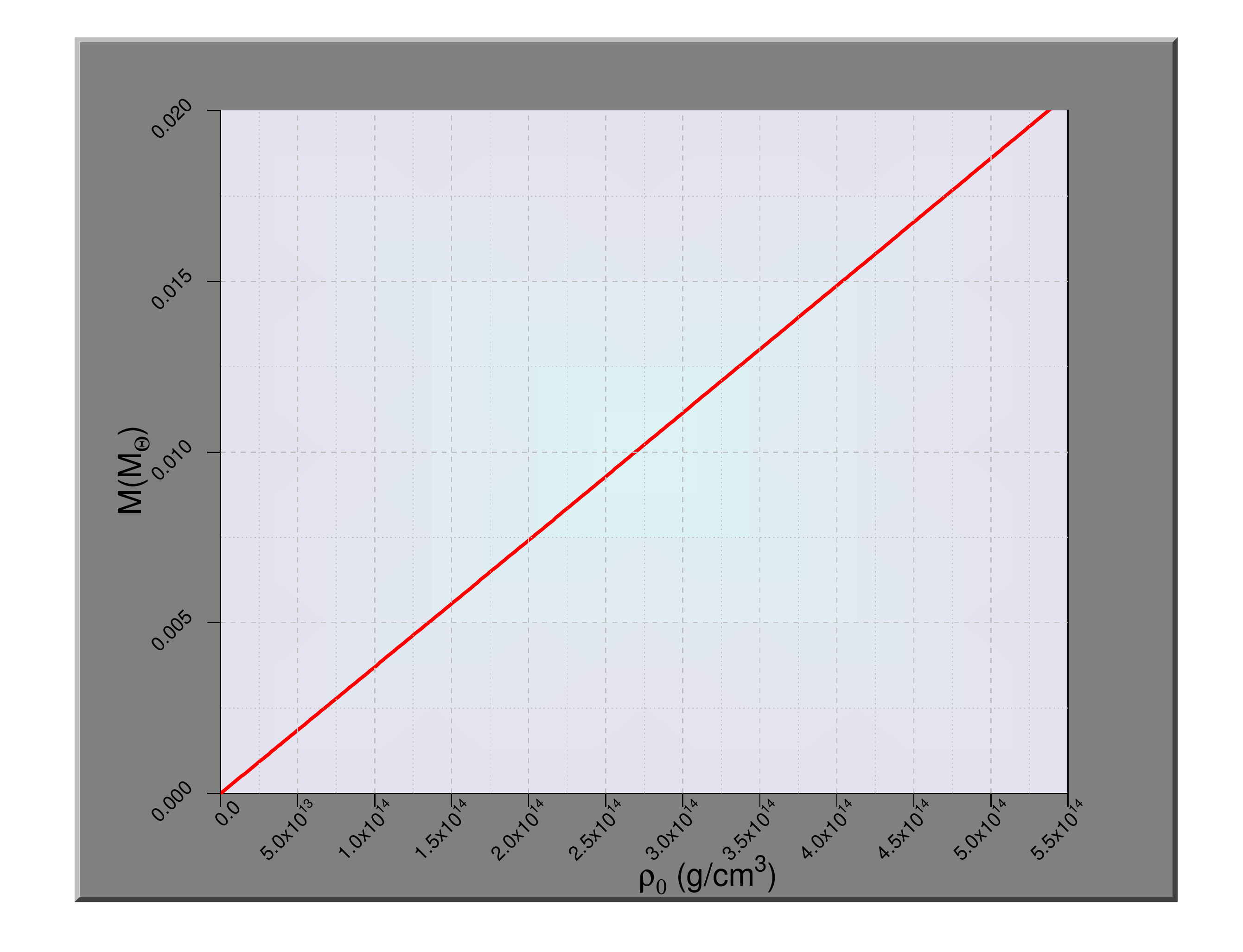} 
				\caption{Behavior of Mass M vs. central density $\rho_0$ for PSR \textquotedblleft B0943+10\textquotedblright.} \label{M}
			\end{minipage}
		\end{center}
	\end{figure}

	We have demonstrated the behavior of adiabatic index $\gamma$ in Fig. (\ref{adia}), which shows the desirable features. The value of $\gamma$ at the center is $4047.839101$, and the graph clearly indicates that with an increase in radius, $\gamma$ increases drastically. 
	
	If we consider a Newtonian sphere to be in stable equilibrium, $\gamma$ must have values strictly greater than $\frac{4}{3}$ throughout the region, and $\gamma= \frac{4}{3}$ is the condition for a neutral equilibrium \cite{adiabatic}. This condition changes for a relativistic isotropic sphere due to regenerative effect of pressure, which makes the sphere more unstable.
	\subsubsection{Harrison-Zeldovich-Novikov Stability Criterion}
	According to Harrison, Zeldovich $\&$ Novikov \cite{harri,zeldo} criterion, for a compact star to be stable, it's mass should increase with rise in central density. Mathematically, $\frac{dM}{d\rho_0}>0$ throughout the stellar configuration. 
	
	Using eqs. (\ref{bcon}) and (\ref{centdens}) we have,
	\begin{equation}
		M=4\pi R^3\frac{\rho_0}{M_1}\Big[K-1+4\pi KR^2\rho_0\frac{ M_2}{M_1}\Big]
	\end{equation}
	where,\\
	$M_1=3(K-1)+8\pi K\rho_0 R^2$ and $M_2=\Big[-\frac{15}{4} \frac{(1-K)^2}{M_1}-\frac{2aM_1}{3(1-K)^2X_1^2(a+bX_1)}+K-\frac{7}{4}\Big]$\\
	
	We observed in fig \ref{M} that mass of the star is positive and increasing with  increase in central density. Thus, we can conclude that the presented model satisfies Harrison-Zeldovich-Novikov criterion. Hence, the model is stable.

\section{Discussion $\&$ Concluding Remarks}
In this paper, a static and spherically symmetric isotropic fluid model has been developed by using well-known Vaidya-Tikekar metric potential.  The solution of this system involves six constants, two of them has been fixed using the junction condition as well as the property of vanishing pressure at the boundary, leaving the remaining constants to be determined from the study of a real compact star. The presented solution satisfies all the physical criteria of a relativistic compact object. A thorough physical analysis has been accomplished for the star PSR B0943+10. All the physical quantities are regular and well-behaved throughout the stellar interior. Energy density and pressure are decreasing functions as we move towards the surface of the star from centre. This model satisfies causality conditions, energy conditions and stability conditions.

The main features of this study can be summarized as follows: 
\begin{enumerate}
	\item To demonstrate that our solution is compatible with a real compact star, we have used the pulsar PSR B0943+10, which has mass $\sim0.02 {M}_\odot$, and radius $\sim2.6$ km \cite{zeldo}. The use of mass and radius as an input parameter helped us to fix four of the constants as $C=-4.01893\times10^{-5} km^{-2}$, $K=0.0119$, $a=0.001$ and $b=0.029$. Furthermore, the use of boundary conditions yields $ A=- 0.177729  $ and $B= 29.657254$.
	\item As shown in Figure \ref{lambnu}, the metric potentials have no singularity either at the center of the pulsar PSR B0943+10 or at the boundary.
	\item Figure \ref{dens} and \ref{press} show that the density and pressure are positive and decreasing towards the center of the pulsar PSR B0943+10. Also, figure \ref{nd} shows that the gradient of the density and pressure are negative; which is a necessary condition for any real star
	configuration. 
	\item  Figure \ref{v} shows that the velocity of sound is strictly less than 1, which is required for any realistic star. Also, it is very clear from figure \ref{ratio} that in our model, pressure at any certain distance from center of the star is dominated by the corresponding density at that point.
	\item  Figure \ref{ec} shows that our model satisfies all the energy conditions, i.e., NEC, WEC, DEC, SEC which supports that the matter destributed inside the star is normal (not any kind of exotic matter).
	\item From figure \ref{red}, it is clear that the red-shift of our model is less than 2.
	\item Figure \ref{fig:fq} shows that the charge function is positive and increasing (from zero at center) towards the surface of pulsar PSR B0943+10. In this model the net charge is effective to balance the mechanism of the force.
	\item As it can be seen in figure \ref{fghe}, our model is in hydrostatic equilibrium under the forces: gravitational force ($F_{g}$), hydrostatic force ($F_{h}$) and electric force ($F_{e}$).
	\item Figure \ref{adia} shows that the resulting model is stable because its adiabatic index is greater than 4/3.
	\item It follows from figure \ref{M} that the mass of the pulsar PSR B0943+10 increases with increase in energy density, i.e., it satisfies the Harrison-Zeldovich-Novikov criterian for stability. 
\end{enumerate}										
Hence, we can conclude that an analytic solution to the Einstein-Maxwell field equations which meets all the requirements of a physically and mathematically admissible solution representing a static spherically symmetric spacetime described by a charged isotropic energy-momentum tensor has been obtained. We have used this solution to study the pulsar PSR B0943+10 as a Vaidya-Tikekar type star. This model could even be useful to elucidate more compact objects apart from PSR B0943+10. \\

\section*{Acknowledgments}
The Authors would like to express their sincere gratitude towards Science and Engineering Research Board (SERB), DST, New Delhi for providing the financial support needed to complete this work. They are also very grateful to the Department of Mathematics, Central University of Jharkhand, Ranchi, India for the necessary support  where the paper has been written and finalized.

\section*{Appendix: Notations used in paper}
To express equations in a simplified manner the following notations have been used in the paper:\\
$A_1= \frac{2}{(1-K)X}\Big(\frac{a+bX}{2(1+X^2)}-\frac{a}{X^2}\Big)$, \ 
$A_2=\frac{a}{b^3}\mathcal{H}(X)+\frac{B}{A}$, \ 
$A_3=\frac{2(a+bX)}{(1-K)X^2}$, \
$A_4=\frac{a(\sec^2t+\cos^2t-2)}{2b^2(a+bX)}$,  \
$A_5=\frac{a+bX}{X}$; \\
$N(r)=-\frac{(A_1 A_2+A_3 A_4) (N_2 A_5+N_5A_2)}{A_2 A_5}+N_1 A_2+ N_2 A_1+N_3 A_4+N_4 A_3$,\\ 
$N_1(r)=-\frac{A_1}{K+Cr^2}+\frac{2}{(1-K)(K+Cr^2)}\Big[\frac{b}{2(1+X^2)}-\frac{X(a+bX)}{(1+X^2)^2}+\frac{2a}{X^3}\Big]$, \
$N_2(r)=\frac{A_4}{(1-K)X}$, \
$N_3(r)=-\frac{1}{K+Cr^2}\Big[A_3+\frac{2a}{K+Cr^2}\Big]$, \ 
$N_4(r)=\frac{1}{(1-K)X}\left[-\frac{bA_4}{a+bX}+\frac{a(\sec^2t-\cos^2t)}{2b(a+bX)^2}\right]$, \ 
$N_5(r)=-\frac{a}{(1-K)X^3}$, \
$N_6(r)=\frac{2(1-K)(5+Cr^2)}{K(1+Cr^2)^3}$, \
$N_7(r)=\frac{1-Cr^2}{K(1+Cr^2)^3}\bigg[\frac{5}{4(1+X^2)}+\frac{2a(1+X^2)}{X^2(a+bX)}+K-\frac{7}{4}\bigg]$, \
$ N_8(r)=\frac{Cr^2}{4K(1-K)(1+Cr^2)^2}\left[\frac{5}{(1+X^2)^2}-\frac{8a}{X^2(a+bX)}+
\frac{4a(1+X^2)(2a+3bX)}{X^4(a+bX)^2}\right]$;\\
$X_1=\sqrt{\frac{K+CR^2}{1-K}}$, \  
$J=\frac{CR^2}{2(1-K)(1+CR^2)}\left[\frac{5}{4}\frac{(1-K)}{(1+CR^2)}+\frac{2a}{{X_1}^2(a+bX_1)}\frac{1+CR^2}{(1-K)}+K-\frac{7}{4}\right]$, \
$A_{11}= \frac{2}{(1-K)X_1}\Big(\frac{a+bX_1}{2(1+{X_1}^2)}-\frac{a}{{X_1}^2}\Big)$, \\
$A_{31}=\frac{2(a+bX_1)}{X_1^2(1-K)}$, \
$A_{41}=\frac{a(sec^2t_1+cos^2t_1-2)}{2b^2(a+bX_1)}$, \ $t_1=\tan^{-1}\sqrt{\frac{bX_1}{a}}$ \ $\&$
$A_{51}=\frac{a+bX_1}{X_1}$. \\

\end{document}